\documentclass[11pt]{article}
\usepackage[latin1]{inputenc}
\usepackage{amsmath}
\usepackage{amsthm}
\usepackage{amsfonts}
\usepackage{amssymb}
\usepackage{graphicx}
\usepackage{cleveref}
\usepackage{epstopdf}
\usepackage{algorithm}
\usepackage[noend]{algpseudocode}
\usepackage{svg}
\usepackage{caption}
\makeatletter
\def\BState{\State\hskip-\ALG@thistlm}
\makeatother

\theoremstyle{definition}

\newtheorem{proposition}{Proposition}

\theoremstyle{remark}
\newtheorem{remark}{Remark}

\newtheorem{assumption}{Assumption}
\date{}

\title{\LARGE \bf Automation Of Road Intersections Using Consensus-based Auction Algorithms}
\author{
	Fabio Molinari\thanks{F. Molinari is with the Technische Universit\"at Berlin, Germany. {\tt\small molinari@control.tu-berlin.de}}
	 \and 
	 J\"org Raisch\thanks{J. Raisch is with the Technische Universit\"at Berlin \& Max-Planck-Institut fur Dynamik Komplexer Technischer Systeme, Germany. {\tt\small raisch@control.tu-berlin.de}\newline\newline{This work was funded by the German Research Foundation (DFG) within their priority project SPP 1914 "Cyber-Physical Networking (CPN)".}\newline{Authors want to thank Juan Antonio Delgado Guerrero for his help with the Python simulation environment.}
	}
}
\begin{document}
	\maketitle
	\begin{abstract}
		This paper investigates a consensus-based auction algorithm in the context of decentralized traffic control. In particular, we study the automation of a road intersection, where a set of vehicles is required to cross without collisions. The crossing order will be negotiated in a decentralized fashion. An on-board model predictive controller (MPC) will compute an optimal trajectory which avoids collisions with higher priority vehicles, thus retaining convex safety constraints. Simulations are then performed in a time-variant traffic environment. 
	\end{abstract}	
	\section{Introduction}
	\label{sec:introduction}
	Stoplights and stop signs are traditionally responsible for governing traffic around a road intersection, by giving distinct priorities to drivers coming from different lanes. Provided that all drivers respect signals and traffic rules, this system prevents collisions. Although safe, it is regarded as inefficient (especially in the case of non-saturated roads) \cite{stoffers1968scheduling}; it can improve if, for a given car density, a solution which increases the average speed of vehicles can be obtained \cite{apostel2005self}. At the same time, many car manufactures are focusing on introducing advanced driver assistance systems (ADAS), which aim to improve traffic efficiency, avoid congestions, and augment throughput \cite{behere2013reference}. If inter-vehicle communication (V2V) is additionally provided, vehicles will be able to adapt their behavior according to the information sent by other traffic participants. A cooperative driving system, therefore, will create an opportunity to get rid of any traditional traffic sign at intersections, since vehicles will be able to team up and agree on how to cross the intersection without collisions. It is possible to refer to \cite{chen2016cooperative} as a comprehensive overview of the challenges that such a technology entails. Originally, central coordinating units were responsible for planning and optimizing oncoming vehicles' trajectories \cite{murgovski2015convex}; decentralized solutions, which are able to negotiate actions and solve conflicts without depending on any central element, have been also investigated \cite{hafner2013cooperative}. They prevent computational complexity from increasing with the number of vehicles. 
	
	We consider model predictive control (MPC) a valid methodology to generate trajectories in such a decentralized context, due to its ability to deal explicitly with constraints and to predict trajectories of other vehicles. In \cite{campos2014cooperative,katriniokdistributed}, a distributed MPC strategy is applied, which guarantees collision-free feasible trajectories, with the aim of optimizing a desired cost function. 
	In \cite{campos2014cooperative}, each agent decides, in a sequentialized fashion, whether to pass the intersection before or after agents with lower priority. 
	Unlike this, \cite{katriniokdistributed} adopts a priority based policy which avoids nested iterations and non-convex safety constraints in the local optimal control problems. 
	Each vehicle has to respect a collision avoidance constraint only with higher priority vehicles. This allows a semi-definite relaxation of the local optimal control problems. 
	However, crossing priorities are still passed as pre-defined lists, which do not evolve with time-varying traffic conditions (i.e., new vehicles approaching the intersection or unexpected variations due to unpredictable events like pedestrians crossing the road).
	
	This paper focuses on a decentralized strategy which allows vehicles to agree on a priority list. Unlike \textit{Remark 2} in \cite{katriniokdistributed}, these lists imply an intersection crossing order. In the context of multi-agent systems, of which road traffic is a valuable example, consensus-based control, whose aim is having agents agree on some quantity, has been recurrently applied \cite{ren2007information}.
	For addressing a task allocation problem among a fleet of robotic agents, a consensus-based strategy has been harnessed in the consensus-based auction algorithm (CBAA) presented in \cite{choi2009consensus}. This makes use of two subsequent moments: first, with a market-based decision strategy, agents negotiate tasks, then, with a consensus-based protocol, the fleet achieves agreement on the winning bid list. We will present a modified version of CBAA algorithm to distributively negotiate crossing priorities. Every vehicle, then, mounts an on-board MPC controller for computing the optimal trajectory, which retains safety constraints only with higher priority vehicles, according to the prior negotiated lists. This makes the local optimal problems convex.
	No restrictions are issued on the maximum number of vehicles approaching the intersection and coming from the same lane. 
	
	The remainder of this paper is organized as follows: the discrete-time vehicle model and road description are presented in \Cref{sec:problem_description}. The modified CBAA is described in \Cref{sec:algorithm} and applied to the problem of automating road intersections in \Cref{sec:priorityLists}. The local optimal control strategy can be found in \Cref{sec:problem_statement}. Simulation results are presented in \Cref{sec:simulation}, while final remarks and future development are finally presented in \Cref{sec:conclusion}.
	
	\subsection{Notation}
	We use $\mathbb{N}$ and $\mathbb{R}$ to denote, respectively, the set of positive integers and the set of real numbers. The set of positive real numbers and nonnegative real numbers are denoted, respectively, by $\mathbb{R}_{>0}$ and $\mathbb{R}_{\geq0}$. Given a matrix $A\in\mathbb{R}^{n\times m}$, with $n,m\in\mathbb{N}$, its element in position $(i,j)$ is denoted by $[A]_{ij}$. Given a vector $v\in\mathbb{R}^n$, its transpose is $v^T$. Two sorting functions are used. The function $\text{sort}:\mathbb{R}_{\geq 0}^n\rightarrow\mathbb{R}_{\geq 0}^n$ arranges vector elements in decreasing order of magnitude, and the function $\text{argsort}:\mathbb{R}_{\geq 0}^n\rightarrow\mathbb{N}^n$ places vector indices in decreasing order of their respective entries' magnitude. The function $\text{argmax}:\mathbb{R}_{\geq 0}^n\rightarrow\mathbb{N}$ returns the index whose correspondent entry exhibits the maximum value in the vector. If the maximum value is contained in more than one entry, only one index, randomly selected among the possible candidates, is returned.
	Given a	set $\mathcal{S}$, its cardinality is $|\mathcal{S}|$.
	
	\section{Problem description}
	\begin{figure}
		\center
		\includegraphics[width=\columnwidth]{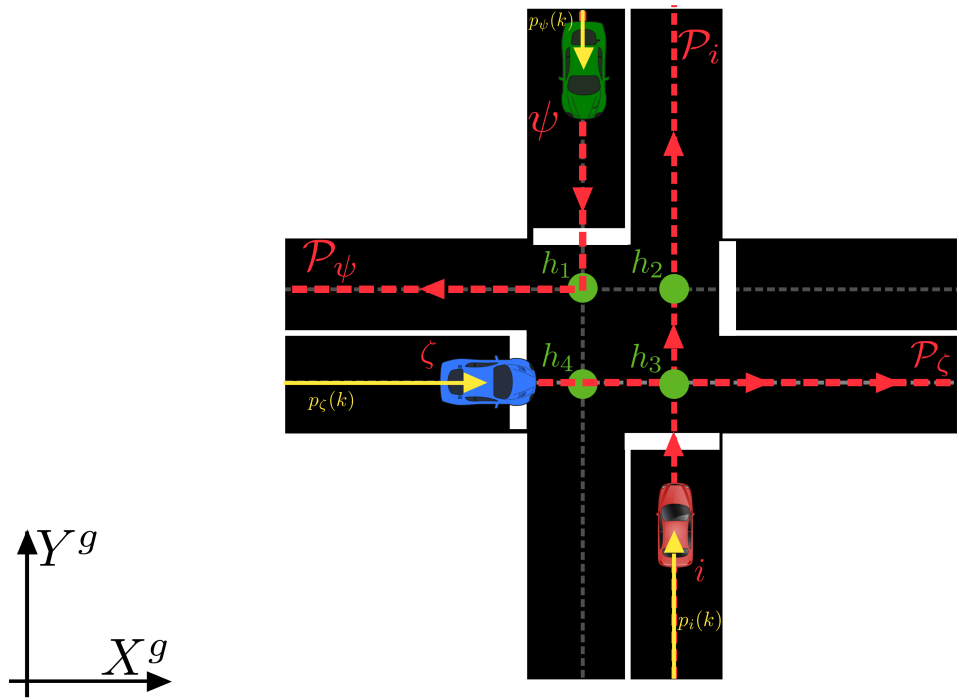}
		\caption{The above situation represents three vehicles approaching a road intersection. Vehicles $i$ and $\zeta$ are going straight along their prescribed paths, $\mathcal{P}_i$ and $\mathcal{P}_\zeta$ respectively, whilst $\psi$ is turning right, coherently with $\mathcal{P}_\psi$. In the general case, the paths corresponding to all the vehicles going straight or turning right describe four possible collision points, highlighted in the figure. }
		\label{fig:road}
	\end{figure}
	\label{sec:problem_description}
	In the following, we analyze the road intersection represented in \Cref{fig:road}. Each road has two lanes, and a right-hand traffic environment is assumed. 
	
	A set $\textbf{N}=\{1,\dots,N\}$ of $N>1$ vehicles approaching the intersection is given. Every vehicle $i\in\textbf{N}$ is provided with an arbitrary path in global coordinates, i.e. $\mathcal{P}_i=(\mathbb{X}_{i}^g,\mathbb{Y}_{i}^g)\subset\mathbb{R}^2$, which can either be straight or right-turning. Acceleration can be varied along these paths. 
	In principle, the present framework could allow also left-turning. However, in this paper, for simplicity and without loss of generality, this case is not considered.
	Furthermore, many delivery companies are discouraging left-turning (\cite{moneyups,carraher2014results}), in order to decrease safety hazards and delays.
	Additionally, overtaking in the vicinity of an intersection is not allowed. 
	As in \cite{murgovski2015convex}, let $\mathbf{x}_i(k)=[p_i(k)\ v_i(k)]^T$ be the state vector of vehicle $i\in\textbf{N}$ along its path $\mathcal{P}_i$, where the longitudinal position and speed along the path, respectively $p_i(k)$ and $v_i(k)$, are sampled at discrete instants $k\in\mathbb{N}$, with a sampling time which equals $T_s\in\mathbb{R}_{>0}$. Then, every vehicle $i\in\textbf{N}$ is represented by the point-mass linear discrete-time system
	\begin{equation}
		\label{eq:linSysVeh_i}
		\mathbf{x}_i(k+1)=A\mathbf{x}_i(k)+Bu_i(k),
	\end{equation}
	with
	\begin{equation}
		\label{eq:sysMatrices}
		A=\left[
			\begin{matrix}
				1 &T_s\\
				0 &1
			\end{matrix}
		\right] ,\ 
		B=\left[
		\begin{matrix}
		0 \\
		T_s 
		\end{matrix}
		\right]		,
	\end{equation}
	where the control signal $u_i(k)$ represents the longitudinal acceleration along the path. Every vehicle $i\in\textbf{N}$ is provided with a map $\mathcal{M}_i:\mathbb{R}\rightarrow\mathcal{P}_i$, which transforms the longitudinal position along $\mathcal{P}_i$ at instant $k\in\mathbb{N}$ to the correspondent position in global coordinates, i.e. $\forall i\in\textbf{N},\  \exists\mathcal{M}_i :$
	\begin{multline}
		\label{eq:glob_coo_trans}
		\mathcal{M}_i(p_i(k))=\mathbf{x}_i^{g}(k)=\\(x^g_i(k)\in\mathbb{X}_{i}^g,\  y^g_i(k)\in\mathbb{Y}_{i}^g)\in\mathcal{P}_i\subset\mathbb{R}^2,
	\end{multline} 
	$\forall k\in\mathbb{N}$.	
	The inverse map $\mathcal{M}_i^{-1}:\mathcal{P}_i\rightarrow\mathbb{R}$ is also defined, which performs a global to local transformation. 
	
	All the possible paths, i.e. $\{\mathcal{P}_i\mid i\in\textbf{N}\}$, identify a set of four shared points, 
	\begin{equation}
		\mathcal{H}=\{\mathbf{h}_j=(h_{j_x},h_{j_y})\in\mathbb{R}^2\mid 1\leq j\leq 4,\ j\in\mathbb{N}\}, 
	\end{equation}	
	as in \Cref{fig:road}. Every vehicle, according to its own path, will be approaching either one (in case it turns right) or two shared points (in case it goes straight). Given a pair of cars $i_1,i_2\in\textbf{N}$, a collision may happen at instant $k\in\mathbb{N}$ if
	\begin{multline}
		\label{eq:collision}
		\text{d}(\mathbf{x}_{i_1}^{g}(k),\mathbf{x}_{i_2}^{g}(k))=\\\sqrt{(x_{i_1}^g(k)-x_{i_2}^g(k))^2+(y_{i_1}^g(k)-y_{i_2}^g(k))^2}\leq d_s,
	\end{multline}
	where $d_s$ is a measure which models cars' dimensionality. 
	Additionally, all cars driving at instant $k\in\mathbb{N}$ in front of $i\in\textbf{N}$ on its path can be grouped into a set, which is represented by $\mathcal{F}_i^k$, defined as:
	\begin{multline}
		\label{eq:def_frontal}
		\mathcal{F}_i^k=\{\zeta\in\textbf{N}\mid \mathbf{x}_\zeta^{g}(k)\in\mathcal{P}_i\ \land \mathcal{M}_i^{-1}(\mathbf{x}_\zeta^{g}(k))>p_i(k) \}.
	\end{multline}
	Accordingly, collisions may occur either between cars driving on the same lane (rear-end collision) or between cars crossing the same shared point and coming from different directions. Therefore, we name $\mathcal{H}$ the set of collision points. 
	The traffic is in full composed of autonomous vehicles, whose vehicle-to-vehicle (V2V) communication graph is assumed to be fully connected. The assumption of a common shared clock amongst vehicles allows us to neglect any communication delay and resulting non-idealities.

	\section{Consensus-based auction algorithm}
	\label{sec:algorithm}
	The algorithm proposed here takes inspiration from a CBAA presented in \cite{choi2009consensus}, and it will be referred to as CBAA-M (consensus-based auction algorithm modified). A set $\mathcal{S}$ of $S>1$ synchronously communicating agents is given. 
	The communication topology can be represented by a directed graph $\mathcal{G}$, with node set $\mathcal{S}$ and edge set $\mathcal{E}\in\mathcal{S}\times\mathcal{S}$. If $(i,j)\in\mathcal{E}$, agent $i\in\mathcal{S}$ will be able to transmit information to agent $j\in\mathcal{S}$. Note that $(i,i)\in\mathcal{S}$, $\forall i\in\mathcal{S}$. 
	The goal is to have agents agree on their order, based on some rule. As in \cite{goldberg2003competitiveness}, agents can compete by placing bids in order to get the highest possible position in the list. Accordingly, each agent $i\in\mathcal{S}$ retains an asset $c_i\in\mathbb{R}_{>0}$, which represents its bid (in the following, without loss of generality, we assume $c_{i_1}\not=c_{i_2}$, $\forall i_1\not=i_2\in\mathcal{S}$). A centralized solution would just sort agents according to their bids. However, without any central element, the following protocol is used, which is composed of two sequential phases, an asynchronous auction process and the consensus algorithm which makes agents converge to an ordered list. Each iteration $\kappa$ of the algorithm consists of a single run of phase 1 (performed asynchronously by all agents) and phase 2, assumed to be run synchronously by the whole set of agents. 
	\begin{algorithm}
		\caption{CBAA-M Phase 1 for agent $i$ at iteration $\kappa$}\label{cbaam1}
		\label{algo:cbaam1}
		\begin{algorithmic}[1]
			\State $\mathbf{v}_i^{0}=\mathbf{0}_S$ , $\mathbf{w}_i^{0}=\mathbf{0}_S$
			\Procedure{Bid($c_i$,\ $\mathbf{v}_i^{\kappa-1}$,\ $\mathbf{w}_i^{\kappa-1}$)}{}
			\State $\mathbf{v}_i^{\kappa}\gets \mathbf{v}_i^{\kappa-1}$
			\State $\mathbf{w}_i^{\kappa}\gets \mathbf{w}_i^{\kappa-1}$
			\State $j\gets 0$
			\BState \emph{loop}:
			\If {$i\not=(\mathbf{v}_{i}^{\kappa})_l$, $\forall l=1\dots S$}
			\If {$c_i>(\mathbf{w}_{i}^{\kappa-1})_j$}
			\State $(\mathbf{v}_{i}^{\kappa})_j \gets i$
			\State $(\mathbf{w}_{i}^{\kappa})_j \gets c_i$
			\EndIf
			\State $j\gets j+1$		
			\State \textbf{goto} \emph{loop}.
			\EndIf
			\State \textbf{close};
			\EndProcedure
		\end{algorithmic}
	\end{algorithm}	
	
	Agent $i\in\mathcal{S}$ keeps two vectors, $\mathbf{v}_i^\kappa$ and $\mathbf{w}_i^\kappa$ (at iteration $\kappa$) of dimension $S$, which will converge to the ordered list of agents' indices and to the ordered list of corresponding bids, respectively. At step $0$, they are initialized as zero vectors of dimension $S$. In the first phase, in \Cref{algo:cbaam1}, each agent $i\in\mathcal{S}$ updates its own vectors asynchronously with all the others, by running an auction algorithm. Accordingly,  if no elements of vector $\mathbf{v}_{i}^{\kappa}$ has value $i$, the agent writes its index $i$ in the vector $\mathbf{v}_{i}^{\kappa}$, directly below the indices of agents who placed higher bids. In the same position, but in $\mathbf{w}_i^\kappa$, the bid $c_i$ is put.

	\begin{algorithm}
		\caption{CBAA-M Phase 2 for agent $i$ at iteration $\kappa$}\label{cbaam2}
		\begin{algorithmic}[1]	
			\label{algo:cbaam2}		
			\State $\text{SEND}$ $(\mathbf{v}_i^{\kappa},\mathbf{w}_i^{\kappa})$ to $j\in\bar{\mathcal{S}}_i=\{j\in\mathcal{S}\mid (i,j)\in\mathcal{E}\}$
			\State $\text{RECEIVE}$ $(\mathbf{v}_h^{\kappa},\mathbf{w}_h^{\kappa})$ from $h\in{\underline{\mathcal{S}}}_i=\{j\in\mathcal{S}\mid (j,i)\in\mathcal{E}\}$
			\Procedure{Update($\mathbf{v}_{h\in\underline{\mathcal{S}}_i}^{\kappa}$, $\mathbf{w}_{h\in\underline{\mathcal{S}}_i}^{\kappa}$)}{}
			\State $(\mathbf{{a}}_{i}^{\kappa})_j \gets \text{argmax}_{h}((\mathbf{w}_{h}^{\kappa})_j)\ \ \ \forall j=1\dots \kappa$
			\State $(\mathbf{v}_{i}^{\kappa})_j \gets (\mathbf{v}_{(\mathbf{{a}}_{i}^{\kappa})_j}^{\kappa})_j\ \ \ \forall j=1\dots \kappa$
			\State $(\mathbf{w}_{i}^{\kappa})_j \gets \max_{h}((\mathbf{w}_{h}^{\kappa})_j)\ \ \ \forall j=1\dots \kappa$
			\EndProcedure
		\end{algorithmic}
	\end{algorithm}
	Then, a second phase (see \Cref{algo:cbaam2}) is needed, in order to make neighbouring agents (in the sense of the underlying communication graph) agree on these two vectors. Each agent $i\in\mathcal{S}$, at each iteration $\kappa$, sends its updated vectors to agents $j\in\bar{\mathcal{S}}_i=\{j\in\mathcal{S}\mid (i,j)\in\mathcal{E}\}$ and receives updated vectors $\mathbf{v}_h^{\kappa}$ and $\mathbf{w}_h^{\kappa}$ from agents $h\in\underline{\mathcal{S}}_i=\{j\in\mathcal{S}\mid (j,i)\in\mathcal{E}\}$.
	
	The expression 
	\begin{equation}
		\text{argmax}_{h}((\mathbf{w}_{h}^{\kappa})_j)
	\end{equation}
	retrieves, among all the agents in $\underline{\mathcal{S}}_i$, the index of the agent knowing who placed the largest bid at iteration $\kappa$ for the position $j$ in the priority list. This agent, whose index is stored in $(\mathbf{{a}}_{i}^{\kappa})_j$, will be asked the winner's index for position $j$, i.e. $(\mathbf{v}_{(\mathbf{{a}}_{i}^{\kappa})_j}^{\kappa})_j$.
	A simple max-consensus protocol is, then, run to retrieve the max bid among neighbours for each priority position $j$, i.e. $\max_{h}((\mathbf{w}_{h}^{\kappa})_j)$.
	
	The following proposition aims at proving the finite-time convergence of the proposed algorithm, provided that the communication graph is fully connected, i.e. $(i,j)\in\mathcal{E}$, $\forall i,j\in\mathcal{S}$.
	\begin{proposition}
		Given $\mathcal{S}$, a set of $S>1$ agents, whose communication topology is represented by a fully connected graph, by applying sequentially \Cref{algo:cbaam1} and \Cref{algo:cbaam2}, agreement on vectors $\mathbf{v}$ and $\mathbf{w}$ is reached in $S$ iterations, i.e.
		\begin{align}
			\mathbf{v}_i^S&=\mathbf{v}_j^S=\mathbf{v}^*=\text{argsort}(\mathbf{c})\\
			\mathbf{w}_i^S&=\mathbf{w}_j^S=\mathbf{w}^*=\text{sort}(\mathbf{c})
		\end{align}
		$\forall i,j\in\mathcal{S}$, where $\mathbf{c}=\{c_i\mid i\in\mathcal{S}\}$.
		
		\begin{proof}
			In a fully connected network, max-consensus is reached in one step \cite{nejad2009max}. Then, at the end of each iteration $\kappa$,
			\begin{equation}
			\label{eq:proof_cbaam}
			\textbf{v}_{i}^\kappa =\textbf{v}_{l}^\kappa,\ \
			\textbf{w}_{i}^\kappa =\textbf{w}_{l}^\kappa,
			\end{equation} 
			$\forall i,l\in\mathcal{S}$, regardless of the outcome of Phase 1. 
			By running the algorithm, at each iteration $\kappa$, $\textbf{v}_i^\kappa$ and $\textbf{w}_i^\kappa$ after Phase 2 will retain the ordered indices of the $\kappa$ agents with the highest bids and their correspondent bids in the first $\kappa$ positions. The last $S-\kappa$ elements will be $0$, as initialized at iteration $\kappa=0$.			
			Convergence is reached when the agent with the lowest bid, e.g. $\underline{i}\in\mathcal{S}$, gets its index and bid stored in position $S$ in $\textbf{v}_{\underline{i}}^\kappa$ and $\textbf{w}_{\underline{i}}^\kappa$, respectively. With the subsequent Phase 2, all the other agents $j$ will get these entries in position $S$ of vectors $\textbf{v}_i^\kappa$, $\textbf{w}_i^\kappa$ (since $c_{\underline{i}}>0$). According to the above analysis for iteration $\kappa$, this will take $\kappa=S$ iterations. 
			\qed
		\end{proof}
	\end{proposition}
	Future work will deal with non fully connected communication topologies, by making use of the results presented in \cite{nejad2009max}, which provides an accurate analysis of convergence properties for max-consensus problems in not necessarily fully connected graphs.
	\section{Priority lists for collision zones}
	\label{sec:priorityLists}
	According to \Cref{sec:problem_description}, each vehicle $i\in\textbf{N}$ will be crossing one or two collision points. Let $G_k$ be a function that associates, at each time step $k\in\mathbb{N}$, every vehicle to the collision points which it is going to cross, i.e. $G_k:\textbf{N}\rightarrow 2^\mathcal{H}$, such that
	\begin{equation}
		G_k(i)=\{ \textbf{h}_j \in \mathcal{H} \mid \textbf{h}_j\in\mathcal{P}_i \land \mathcal{M}_i^{-1}(\textbf{h}_j)>p_i(k) \}.
	\end{equation}
	Define $H_k:\mathcal{H}\rightarrow2^\textbf{N}$ as the dual function, which, for every collision point, gives a list of vehicles going to cross it, i.e.
	\begin{equation}
	H_k(\textbf{h}_j)=\{ i\in\mathbf{N} \mid \textbf{h}_j\in\mathcal{P}_i \land \mathcal{M}_i^{-1}(\textbf{h}_j)>p_i(k) \}.
	\end{equation}
	Obviously, $i\in H_k(G_k(i))$, $\forall i\in\mathcal{S}$, and $\textbf{h}_j\in G_k(H_k(\textbf{h}_j))$, $\forall \textbf{h}_j\in\mathcal{H}$.
	The idea is to use the algorithm presented in \Cref{sec:algorithm} to make agents agree on a crossing priority list for every collision point (in the following also referred to as crossing order list). Namely, $\forall \textbf{h}_j\in\mathcal{H}$, $H_k(\textbf{h}_j)$ is a set of vehicles (agents) which should be agreeing on a priority list at instant $k$, by executing the algorithm presented in \Cref{sec:algorithm}.
	\begin{assumption}
		The execution time of the algorithm in \Cref{sec:algorithm}, namely $T_{a}$, is assumed to be negligible with regard to the sampling time of the system, i.e. $T_{a}\ll T_s$.
	\end{assumption}
	The bid placed by each agent $i\in H_k(\textbf{h}_j)$ at sampling instant $k\in\mathbb{N}$ depends on its dynamics and on the absolute distance to the collision point $\textbf{h}_j$, and it will be
	\begin{equation}
		\label{eq:cost_def}
		c_i^{(j)}(k)=\frac{p_{v}v_i(k)+p_{d}}{\text{d}_{ij}(k)+\epsilon}\ ,
	\end{equation}
	$\forall i \in H_k(\textbf{h}_j)$, $p_v\in\mathbb{R}_{>0}$, $p_d\in\mathbb{R}_{>0}$, $\epsilon\in\mathbb{R}_{>0}$, $v_i(k)$ the longitudinal speed of vehicle $i$, and
	\begin{multline}
		\label{eq:def_dist_veh_col}
		\text{d}_{ij}(k)=\text{d}(\mathbf{x}_{i}^{g}(k),\textbf{h}_j)=\\
		\sqrt{(x_i^g(k)-h_{j_x})^2+(y_i^g(k)-h_{j_y})^2}\
	\end{multline}
	the distance between vehicle $i$ and its future collision point $\textbf{h}_j$.
	\begin{figure}
		\includegraphics[width=\columnwidth]{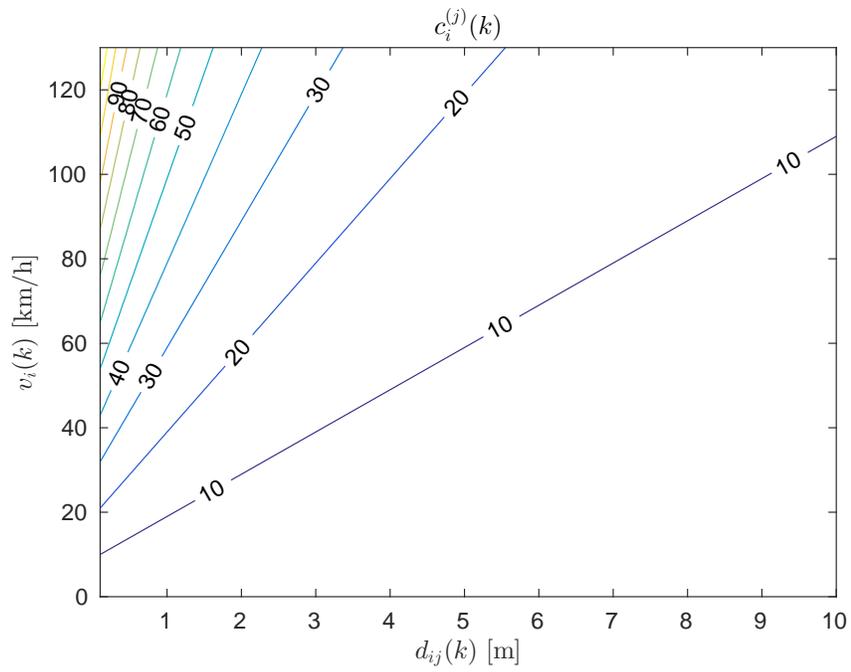}
		\caption{
			Given constant parameters ($p_d=p_v=1$ and $\epsilon=0.1$), the cost $c_i^{(j)}(k)$ is represented as function of $d_{ij}(k)$ and $v_i(k)$.
		} 
		\label{fig:ci}
	\end{figure}
	In Fig. \ref{fig:ci}, it is shown how $c_i^{(j)}(k)$ varies with speed, $v_i(k)$, and distance, $\text{d}_{ij}(k)$, if the tuning parameters $p_v$, $p_d$, and $\epsilon$ are (arbitrarily) chosen as $1$, $1$, and $\frac{1}{10}$. Accordingly, if two vehicles have the same speed, the one closer to the intersection will have higher priority. Conversely, if two vehicles have the same distance from the collision point, the faster one will be placing a higher bid. 	
	Parameter $\epsilon$ prevents numerical errors when the distance to the collision point tends to $0$, while $p_v$ and $p_d$ can be regarded as the weight coefficients of the time-to-collision and of the distance to the collision point, respectively. This can be seen by rewriting (\ref{eq:cost_def}) as	
	\begin{equation}
	c_i^{(j)}(k)=p_v\left( \frac{v_i(k)}{\text{d}_{ij}(k)+\epsilon } \right) + p_d \left( \frac{1}{\text{d}_{ij}(k)+\epsilon } \right).
	\end{equation}
	
	Consequently, at each instant $k\in\mathbb{N}$, each vehicle $i\in \textbf{N}$ places a bid, $c_i^{(j)}(k)$, for each collision point $\textbf{h}_j\in G_k(i)\subset\mathcal{H}$ it is going to cross. 
	Each vehicle $i$ is then assigned a set of ordered lists per sampling instant, i.e. $V_k^i=\{(\mathbf{v}^*)^{(j)}_k\mid \textbf{h}_j\in G_k(i)\}$. Note that, because left turns are disallowed, $|V_k^i|\leq 2$. However, according to the strategy introduced in \Cref{sec:problem_description}, each agent only needs to know the set of agents that have higher priority at the possible collision points. This set is given by
	\begin{equation}
		\mathcal{L}_k^i=\{ \textbf{v}_l \mid \textbf{v}\in V_k^i,\ l<m,\ \textbf{v}_m=i \}. 
	\end{equation}	
	Each vehicle $i\in\textbf{N}$, then, is required to avoid collisions with any vehicle $\zeta\in\mathcal{L}_k^i$. Accordingly, any agent $l\in\textbf{N}\setminus \{i\}$, such that $\mathcal{L}_k^l\ni i$, will have to take care of avoiding collisions with $i$.
	\begin{remark}			
		If two cars approach the intersection from the same lane, the frontal car's bid has to be higher, thus preventing a contradiction between the priority list and the requirement that forbids overtaking. \Cref{prop:coherence_frontal_priority} shows that this assumption is realistic, since it translates to constraining rear car's speed as function of the frontal's, which is consistent with the rear-end collision avoidance which will be presented in \Cref{sec:problem_statement}.
	\end{remark}
	\begin{proposition}
		\label{prop:coherence_frontal_priority}
		Given $i,\zeta\in{H}_k(\textbf{h}_j) : \zeta\in\mathcal{F}_k^i$, the following holds
		\begin{equation}
			\label{eq:suff_cond_coherence_list}
			v_i(k)<v_\zeta(k)+\frac{d_s(v_\zeta(k)+\frac{p_d}{p_v})}{\text{d}_{\zeta j}(k)+\epsilon}\implies c_i^{(j)}(k)<c_\zeta^{(j)}(k).
		\end{equation}
		\begin{proof}
			From (\ref{eq:cost_def}), and given $i,\zeta\in\textbf{N}$, such that $\textbf{h}_j\in G_k(i)\cap G_k(\zeta)$ and $\zeta\in\mathcal{F}_k^i$, it follows that $c_i^{(j)}<c_\zeta^{(j)}$ if and only if
			\begin{equation}
				  v_i(k)<v_\zeta(k)+\frac{(\text{d}_{\zeta j}(k)-\text{d}_{ij}(k))(v_\zeta(k)+\frac{p_d}{p_v})}{\text{d}_{\zeta j}(k)+\epsilon}.
			\end{equation}
			However, since $i$ and $\zeta$ are on the same lane, 
			\begin{equation}
				\text{d}_{\zeta j}(k)-\text{d}_{i j}(k)=\text{d}(\mathbf{x}_{i}^{g}(k),\mathbf{x}_{\zeta}^{g}(k)),
			\end{equation}
			which, due to the (\ref{eq:collision}), is constrained to be bigger than $d_s$ in order to avoid collision, thus implying the sufficient condition (\ref{eq:suff_cond_coherence_list}). 
			\qed
		\end{proof}
	\end{proposition}

	\section{Problem statement}
	\label{sec:problem_statement}
	Each vehicle $i\in\textbf{N}$ has a local on-board controller, which outputs, at every sampling instant $k\in\mathbb{N}$, the appropriate acceleration $u_i(k)$. 
	Model predictive control (MPC) is widely seen as an appropriate technique to address control problems involving explicit constraints and that need, in particular, to embody foreseen trajectories of conflicting vehicles. The MPC prediction horizon is referred to as $T_H$. We define $\mathcal{I}_k^i\subseteq \textbf{N}\setminus\{i\}$ as the set of vehicles which $i\in\textbf{N}$ has to be aware of. This set can be seen as the union of $i$'s frontal and higher priority vehicles at instant $k\in\mathbb{N}$, i.e.
	\begin{equation}
		\mathcal{I}_k^i=\mathcal{F}_k^i\cup\mathcal{L}_k^i.
	\end{equation}
	
	
	\subsection{Prediction model}
	As in \cite{kural2010model}, the prediction is realized based on the information of the current states, which	are used to extrapolate the future behavior of the model until	the end of the horizon. 
	Predicted state and input of $i\in\textbf{N}$ itself for $t\leq T_H$ are referred to as $\tilde{\mathbf{x}}_i(t)=[\tilde{p}_i(t)\ \tilde{v}_i(t)]^T$ and $\tilde{u}_i(t)$, respectively, and evolve along the horizon as
	\begin{equation}
		\label{eq:MPC_otherdyn}
		\tilde{\mathbf{x}}_i(t+1)=A\tilde{\mathbf{x}}_i(t)+B\tilde{u}_i(t).
	\end{equation}
	The local position can be mapped to global coordinates, as in (\ref{eq:glob_coo_trans}), whose notation will be $\tilde{\mathbf{x}}_i^{g}(t)$, $\forall t\leq T_H$. 	
	The initial state is constrained as $\tilde{\mathbf{x}}_i(0)=\mathbf{x}_i(k)$. 
	
	Also, the dynamics of any other vehicle of interest for the problem, $\zeta\in\mathcal{I}_k^i$, is predicted along the finite-time horizon, under a constant acceleration assumption, i.e.
	\begin{align}
		\label{eq:MPC_mydyn}
		\tilde{\mathbf{x}}_\zeta(t+1)=A\tilde{\mathbf{x}}_\zeta(t)+B\tilde{u}_\zeta,
	\end{align}
	$\forall t\leq T_H$, where $\tilde{u}_\zeta=u_\zeta(k)$ and $\tilde{\mathbf{x}}_\zeta(0)=\mathbf{x}_\zeta(k)$ (every vehicle is required to broadcast its current state, input, and map to the other traffic participants). Similar to $i$, global coordinates are obtained also for $\zeta$, by mapping with $\mathcal{M}_\zeta$, and are referred to as $\tilde{\mathbf{x}}_\zeta^{g}(t)$, $\forall t\leq T_H$. 
	\subsection{Obstacle avoidance}
	The inter-vehicular distance between the controlled vehicle $i\in\textbf{N}$, whose speed is $\tilde{v}_{i}$, and a target one $\zeta\in\mathcal{I}_k^i$ should be constrained to be bigger than a safety distance measure, defined in \cite{martinez2007safe} as 
	\begin{equation}
		\label{eq:definition_dsafe}
		d_{safe}= \lambda_2\tilde{v}_{i} + \lambda_3,
	\end{equation}
	which is often called \textit{Constant Time-headway rule}, where $\lambda_2$ is referred to as the Time-headway \cite{chien1992automatic}, and $\lambda_3$ can be seen as $d_s$, the physical minimum distance introduced in (\ref{eq:collision}). As in \cite{kerrigan2000soft}, a slack variable $\delta(t)$, $\forall t\in[0,T_H]$, is introduced in order to get a soft-constraint for the obstacle avoidance. This prevents a potential infeasibility (and therefore a subsequent absence of a feasible control action) in case $i$ keeps a distance to any $\zeta\in\mathcal{I}_k^i$ which is smaller than the desired one. All the slack variables $\delta(t)$ will then be weighted in the cost function, in order to discourage any attempt of keeping a smaller safety distance. The set of constraints can therefore be represented by
	\begin{align}
		\text{d}(\tilde{\mathbf{x}}_{i}^{g}(t),\tilde{\mathbf{x}}_{\zeta}^{g}(t))&\geq \lambda_2\tilde{v}_i(t)+\lambda_3+\delta(t),\ \ \ 
		\forall \zeta\in\mathcal{I}_k^i \label{eq:obst_avoid_constr_nonconv}\\
		\delta(t)&\in [-\bar{\lambda}_2\tilde{v}_i(t),\ \bar{\delta}]
		\label{eq:MPC_safety3}
	\end{align}
	$\forall t\in[0,H]$, where $0\leq\bar{\lambda}_2<\lambda_2$ and $\bar{\delta}\in\mathbb{R}_{>0}$ an upper bound. By doing this, the time-headway can be scaled down to $(\lambda_2-\bar{\lambda}_2)$, which is greater than $0$. However, any attempt of diminishing the time-headway below $\lambda_2$ will be highly discouraged in the cost function. Since (\ref{eq:obst_avoid_constr_nonconv}) is a collection of non-convex constraints, solving the local optimal control problem might be a challenging task. A relaxation of (\ref{eq:obst_avoid_constr_nonconv}) becomes, then, desirable. Accordingly, the following propositions are stated.
	\begin{remark}
		In the following, $\mathcal{F}_t^i$ represents the set of vehicles that, at prediction time $t$, are in front of agent $i$ according to that agent's prediction model. According to the definition of $\mathcal{F}_k^i$ (\ref{eq:def_frontal}), the condition $\zeta\in\mathcal{F}_t^i$ may seem to introduce a non-convexity (it depends on $p_i(t)$). However, since any vehicle $\zeta\in\mathcal{I}_k^i$ retains higher priority than $i$ (it crosses first), the condition $\zeta\in\mathcal{F}_t^i$ is equivalent to $\tilde{\mathbf{x}}_{\zeta}^{g}(t)\in\mathcal{P}_i$. 
	\end{remark}
	\begin{proposition}[Rear-end collision]
		Given (\ref{eq:obst_avoid_constr_nonconv}), if $\zeta\in\mathcal{F}_t^i$, then
		\begin{multline}
			\label{eq:diff_same_lane}
			\text{d}(\tilde{\mathbf{x}}_{i}^{g}(t),\tilde{\mathbf{x}}_{\zeta}^{g}(t))=\text{d}_l(\tilde{\mathbf{x}}_{i}^{g}(t),\tilde{\mathbf{x}}_{\zeta}^{g}(t))\\
			:=\mathcal{M}_i^{-1}(\tilde{\mathbf{x}}_{\zeta}^{g}(t))-\tilde{{x}}_{i}(t).
		\end{multline}
		
		\begin{proof}
			If $\zeta\in\mathcal{F}_t^i$, then $\tilde{\mathbf{x}}_{\zeta}^{g}(t)\in\mathcal{P}_i$, therefore it is possible to inversely map $\zeta$'s global coordinates along $\mathcal{P}_i$, and express the distance along this path. Since $\zeta$ is constrained to be in front of $i$ (and the framework does not allow overtaking), $\mathcal{M}_i^{-1}(\tilde{\mathbf{x}}_{\zeta}^{g}(t))-\tilde{{x}}_{i}(t)$ exists and is positive.
			\qed
		\end{proof}
	\end{proposition}
	\begin{remark}
		\label{rem:distance_ag_colp}
		Equation (\ref{eq:diff_same_lane}) can be used also for distances between vehicles and their respective future collision points (\ref{eq:def_dist_veh_col}), which lie, by definition, in front of them. It is possible to show that 
		\begin{equation}
			\label{eq:equiv_dist_coll_veh}
			\text{d}_{ij}(t)=\text{d}_{ij}^l(t):=\mathcal{M}_i^{-1}({\textbf{h}_{j}})-\tilde{{x}}_{i}(t),
		\end{equation}
		for $\textbf{h}_j\in G_k(i)$. In case, at prediction step $t$, vehicle $i$ has already visited the collision point $\textbf{h}_j$, $\text{d}_{ij}^l(t)< 0$ (and (\ref{eq:equiv_dist_coll_veh}) is no longer valid).
	\end{remark}
	\begin{proposition}[Higher priority precedence]
		The condition (\ref{eq:obst_avoid_constr_nonconv}), for $\zeta\in\mathcal{L}_k^i\setminus\mathcal{F}_k^i$, holds true if
		\begin{equation}
			\label{eq:higher_notfrontal_const}
			\text{d}_{ij}^l(t)\geq \lambda_2\tilde{v}_i(t)+\lambda_3+\delta(t),\ \ \ \forall t:\text{d}_{\zeta j}^l(t)\geq 0
		\end{equation}
		for $\textbf{h}_j\in G_k(i)\cap G_k(\zeta)$.
		
		\begin{proof}
			Unlike \cite{katriniokdistributed}, the negotiated priority list implies a crossing order. So, if $\zeta\in\mathcal{L}_k^i$, $\zeta$ will cross $\textbf{h}_j\in G_k(i)\cap G_k(\zeta)$ before than $i$. 			
			For $i$ and $\zeta$ coming from different directions (i.e. $\zeta\in\mathcal{L}_k^i\setminus\mathcal{F}_k^i$), it is easy to visualize in \Cref{fig:road} that their paths form in $\textbf{h}_j$ a right angle, thus the Pythagorean theorem implies that
			\begin{equation}
				\text{d}^2(\mathbf{x}_{i}^{g}(t),\mathbf{x}_{\zeta}^{g}(t))=(d_{ij}(t))^2+(d_{\zeta j}(t))^2.
			\end{equation}
			The following is easily provable, for $\bar{d}\geq 0$: 
			\begin{equation}
				d_{ij}(t)\geq \bar{d} \land d_{\zeta j}(t)\geq 0\implies \text{d}(\mathbf{x}_{i}^{g}(t),\mathbf{x}_{\zeta}^{g}(t))\geq \bar{d}.
			\end{equation}
			Since $\textbf{h}_j\in G_k(i)\cap G_k(\zeta)$, then $d_{\zeta j}(t)=d^l_{\zeta j}(t)$ and $d_{ij}(t)=d^l_{ij}(t)$ (see \Cref{rem:distance_ag_colp}). 
			So, the (\ref{eq:higher_notfrontal_const}) comes as a consequence for $\bar{d}=\lambda_2\tilde{v}_i(t)+\lambda_3+\delta(t)$ and states that as long as $\zeta$ has not crossed $\textbf{h}_j$, $i$ has to keep a safety distance to $\textbf{h}_j$.
			\qed
		\end{proof}
	\end{proposition}
	Accordingly, we can rewrite (\ref{eq:obst_avoid_constr_nonconv}) in an equivalent convex form, as
	\begin{equation}
		\label{eq:MPC_safety1}
		\tilde{{x}}_{\zeta}(t)-\tilde{{x}}_{i}(t)\geq \lambda_2\tilde{v}_i(t)+\lambda_3+\delta(t),\ \ \ \forall \zeta\in\mathcal{F}_t^i,
	\end{equation}
	and
	\begin{multline}
		\label{eq:MPC_safety2}
		\mathcal{M}_i^{-1}({h_{j}})-\tilde{{x}}_{i}(t)\geq \lambda_2\tilde{v}_i(t)+\lambda_3+\delta(t),\\ \forall \zeta\in\mathcal{L}_k^i\setminus\mathcal{F}_t^i:\mathcal{M}_\zeta^{-1}({h_{j}})-\tilde{{x}}_{\zeta}(t)\geq0,
	\end{multline}
	$\forall t\in[0,T_H]$, where $\textbf{h}_j\in G_k(i)\cap G_k(\zeta)$.
	
	\subsection{Input and state sets}
	Due to mechanical and legal limitation, both speed and acceleration are constrained to allowed sets. In particular, $\underline{a}_i\in\mathbb{R}$ refers to the minimum acceleration, due to braking capability of vehicle $i\in\textbf{N}$. Its maximum acceleration, due to traction effectiveness, is represented by $\bar{a}_i\in\mathbb{R}$. Concerning the speed, without any loss of generality in our current example, we assume that $i\in\textbf{N}$ cannot go in reverse, therefore $\underline{v}_i=0$; on the other hand, due to either mechanical limits or legal speed limitations, we define $\bar{v}_i\in\mathbb{R}$ as its maximum speed. Accordingly, these constraints will be applied in the MPC to car $i\in\textbf{N}$ and to vehicles in the set $\mathcal{I}_k^i$, i.e.
	\begin{align}
		\tilde{u}_i(t)&\in[\underline{a}_i, \bar{a}_i]\label{eq:MPC_bound1}\\
		\tilde{v}_i(t)&\in[\underline{v}_i, \bar{v}_i]\\
		\tilde{v}_\zeta(t)&\in[\underline{v}_\zeta, \bar{v}_\zeta],\ \ \ \ \forall \zeta\in\mathcal{I}_k^i,\label{eq:MPC_bound2}
	\end{align}
	$\forall t\in[0,T_H]$.
	\subsection{Performance measure}
	Controller performance is evaluated by the following convex cost function
	\begin{equation}
		\label{eq:cost_fun}
		J=\sum\limits_{t=0}^{T_H }
		q (\tilde{v}_i(t)-v_{r_i})^2
		+ r\tilde{u}_i^2(t) 
		+ \omega\delta(t),
	\end{equation}
	where $q,r\in\mathbb{R}_{>0}$, $\omega\in\mathbb{R}_{<0}$ and $v_{r_i}\in\mathbb{R}_{>0}$. As in \cite{katriniokdistributed}, each vehicle $i\in\textbf{N}$ is provided with a reference speed $v_{r_i}$ coming from a higher level trajectory planner. By weighting $\tilde{u}_i(t)$, the controller punishes accelerations and decelerations, in order to guarantee smoother trajectories thus improving comfort, as in \cite{bichi2010stochastic}. Slack variables $\delta(t)$ are negatively weighted, so that the controller, by minimizing $J$, will attempt to increase $\delta(t)$; this means that larger safety distances will be preferred, according to the magnitude of $\omega$. 
	\subsection{Optimal problem formulation}
	
	Accordingly, the optimal control problem solved by each vehicle $i\in\textbf{N}$ at each sampling instant $k\in\mathbb{N}$ can be formulated as
	\begin{equation}
		\begin{split}
			\min\limits_{\tilde{u}_i(0)\dots \tilde{u}_i(T_H)} &\text{cost function } (\ref{eq:cost_fun}) \\
			\text{s.t}\ \ \ &\text{system dynamics }(\ref{eq:MPC_otherdyn},\ref{eq:MPC_mydyn})\\
			&\text{safety constraints }(\ref{eq:MPC_safety1},\ref{eq:MPC_safety2},\ref{eq:MPC_safety3})\\
			&\text{input and state constraints }(\ref{eq:MPC_bound1} - \ref{eq:MPC_bound2}),
		\end{split}
	\end{equation}
	where constraints are issued $\forall t\in[0,T_H]$. Coherently with the purpose of the MPC strategy, the first optimal computed control input will be injected into the system, so that (\ref{eq:linSysVeh_i}) is executed $\forall k\in\mathbb{N}$ with $u_i(k)=\tilde{u}_i(0)$. 
	\begin{assumption}
		The computation time for solving the optimal problem, $T_{opt}$, is assumed to be negligible with regards to the sampling time of the system, i.e. $T_{opt}\ll T_s$.
	\end{assumption}	
	\begin{table}[t]		 
		\caption{Problem data. }
		\label{tab:problemData}
		\begin{tabular}{c}
			\hline\\
			$L_w=3.5m$, $D_w=30m$, $T_s=0.03 s$, $p_c=1$, $p_d=1$, $\epsilon=0.1$,\\
			$\lambda_2=0.1s$, $\bar{\lambda}_2=0s$, $\lambda_3=3.5m$, $\underline{v}_i=\underline{v}_\zeta=0$,  $\bar{v}_i=\bar{v}_\zeta=130km/h$,\\
			$\underline{a}_i=-9m/s^2$, $\bar{a}_i=5m/s^2$, $q=1$, $r=0.01$, $\omega=-0.1$\\
			$\mu_{ref}=45km/h$, $\sigma_{ref}=\sqrt{5}km/h$
			\\
			\hline
			\\
		\end{tabular}\\
		These parameters are of use in the simulations throughout the paper; also the intersection geometry is taken into account; the width of a single lane is $L_w$, while the length of each road forming the intersection is $D_w$.
	\end{table}
	\section{Simulation}
	\label{sec:simulation}	
	In the following, we evaluate the impact on traffic of the above presented control strategy both from a microscopic and a macroscopic point of view, as defined in \cite{may1990traffic}. 	
			
	\begin{figure}[h]
		\center
		\includegraphics[width=.75\columnwidth]{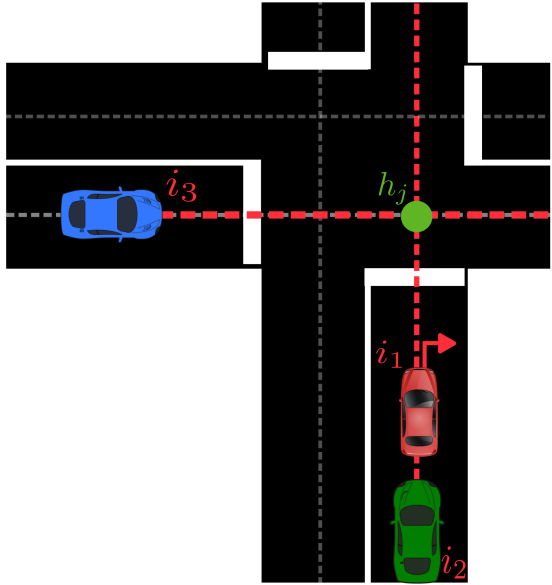}
		\caption{Example scenario with reference (and initial) vehicle speeds $v_{r_{i_1}}=51km/h$, $v_{r_{i_2}}=44km/h$, and $v_{r_{i_3}}=53km/h$ and initial distances from the collision point $\textbf{h}_j$ are $d_{i_1j}(0)=6m$, $d_{i_2j}(0)=14m$, and $d_{i_3j}(0)=11.5m$.
		}
		\label{fig:road_sim}
	\end{figure}	
	
	\begin{figure}[h]
		\center
		\includegraphics[width=\columnwidth]{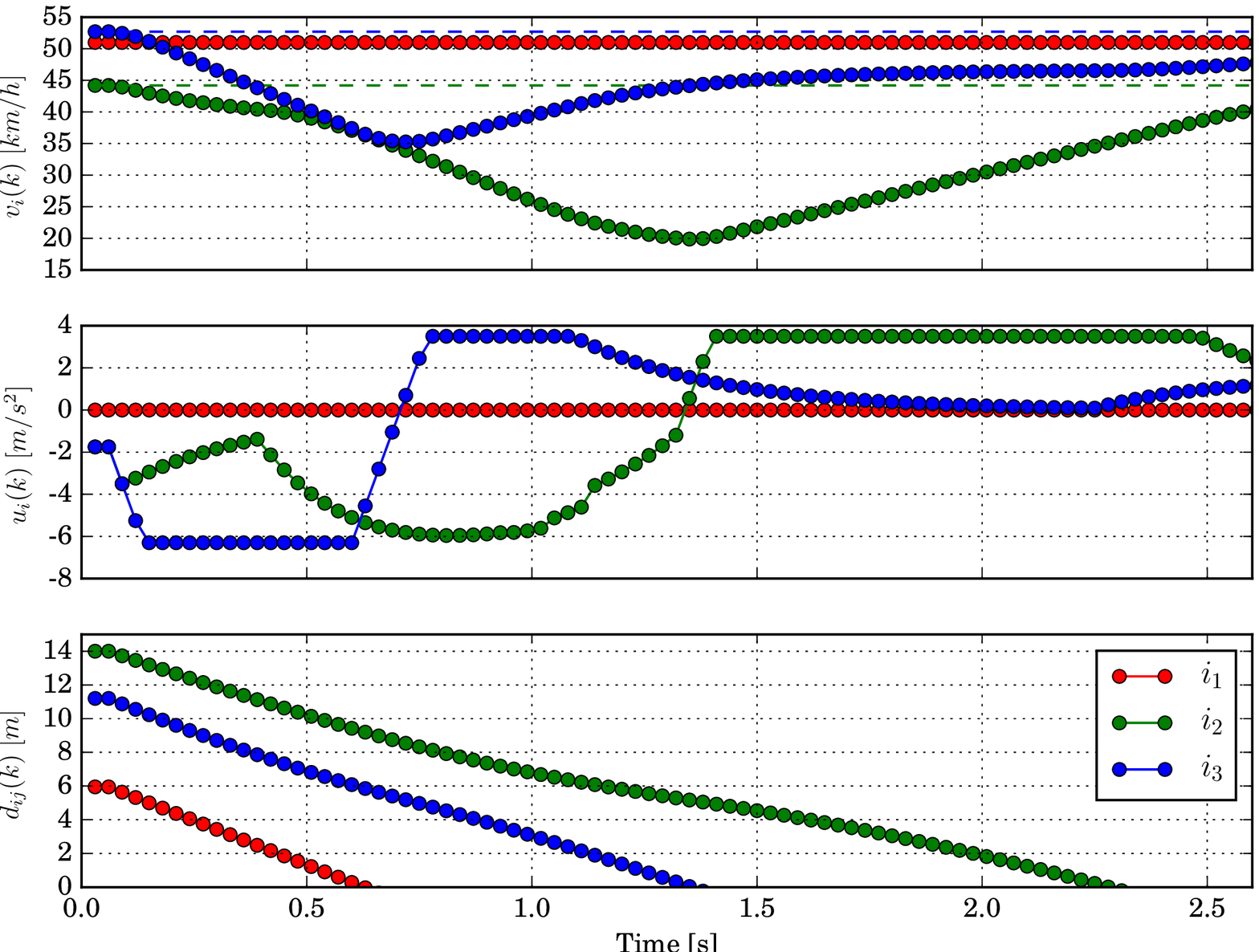}
		\caption{Simulation results for $T_H=100$. Top: speeds as functions of time. Middle: accelerations as functions of time. Bottom: distances to $\textbf{h}_j$ as functions of time.
			}
		\label{fig:scenario_plot}
	\end{figure}
	
	First, the scenario indicated in Fig. \ref{fig:road_sim} is simulated; three vehicles are crossing the intersection, two of whom ($i_1$, $i_2$) coming from the same lane and a third vehicle ($i_3$) intersecting their paths. Both $i_2$ and $i_3$ want to go straight along their respective paths, whilst $i_1$ has to turn right. Vehicle $i_3$ is the fastest (and the one retaining the highest reference speed); nonetheless vehicle $i_1$ places the highest bid according to (\ref{eq:cost_def}), whilst $i_3$ places the second largest. Coherently with \Cref{prop:coherence_frontal_priority}, $i_2$'s bid is smaller than $i_1$'s bid, since $i_1$ is $i_2$'s frontal car. 
	The simulation outcome is plotted in Fig. \ref{fig:scenario_plot}, where speeds, computed accelerations, and distances to the collision point $\textbf{h}_j$ are plotted as functions of time. In order to avoid rear-end collisions, $i_2$ has to proceed slower than its reference speed, until both $i_1$ lies on its path ($t=0.63s$) and $i_3$, which retains a higher priority, has passed the intersection ($t=1.36s$). Vehicle $i_1$ tracks its own reference speed, since it retains the highest bid and, therefore, it has no vehicles to take care of. Conversely, $i_3$ has to slow down, in order to keep a safety distance from $i_1$, the auction winner, which becomes, then, its frontal vehicle. 
	
	\begin{figure}
			\includegraphics[width=\columnwidth]{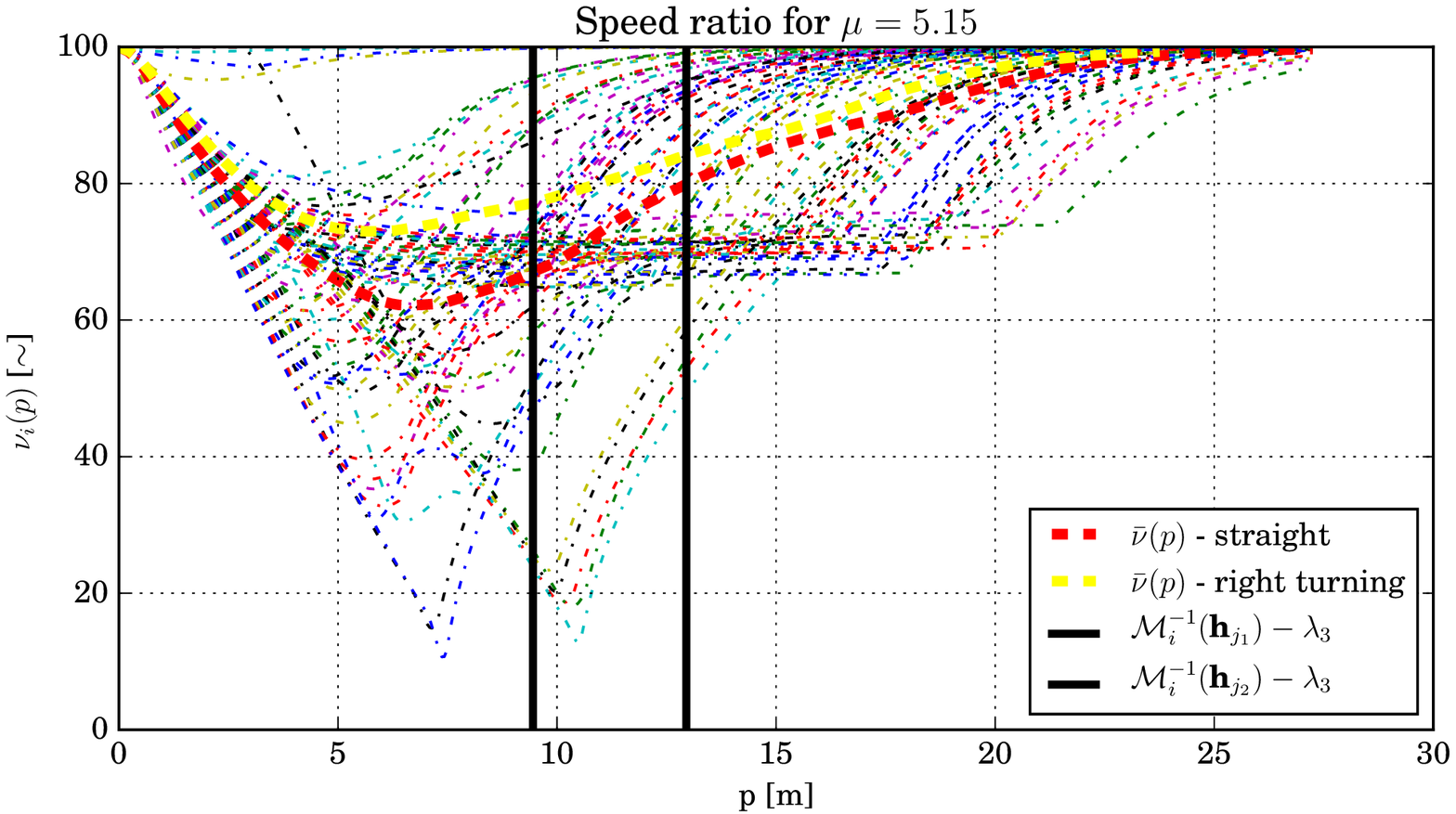}
			\caption{Simulation for $T=90s$, with $H=50$. Each vehicle $i\in\mathbf{N}$ either goes straight (crossing $\textbf{h}_{j_1}$ and $\textbf{h}_{j_2}$) or turns right (crossing only $\textbf{h}_{j_1}$).
				}
			\label{fig:speed_ratio_MD}
	\end{figure}	
	\begin{figure}
			\includegraphics[width=\columnwidth]{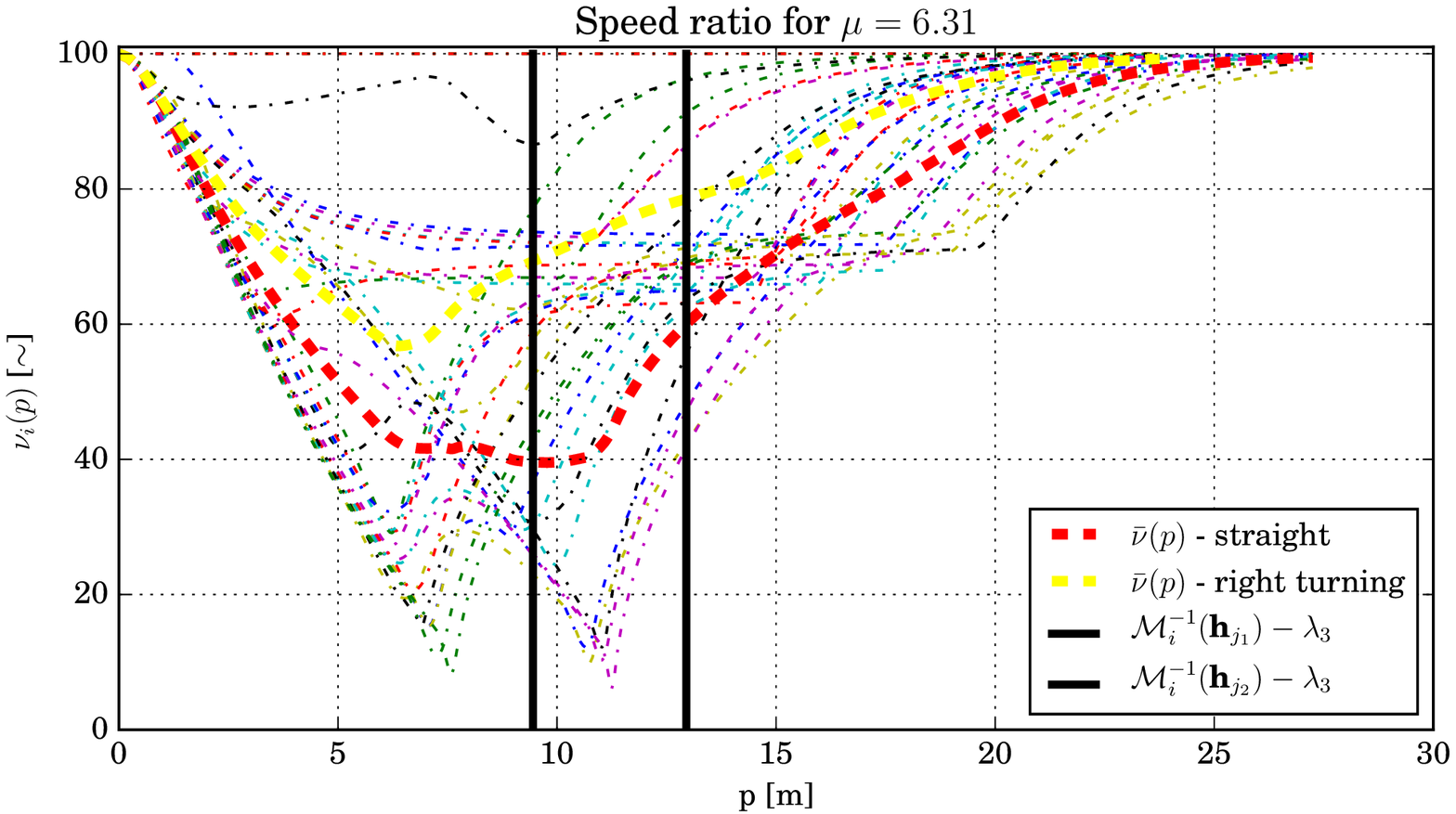}
			\caption{Simulation is run for $T=90s$, with $H=50$.}
			\label{fig:speed_ratio_HD}
	\end{figure}	
	
	The macroscopic impact on traffic is also studied, by simulating a time-varying traffic flow crossing the intersection. A new vehicle enters the simulation environment according to a randomized choice (provided that safety distance is kept with the future frontal vehicle). Vehicles' reference speeds are randomly generated, i.e. $v_{r_i}\sim \mathcal{N}(\mu_{ref},\sigma^2_{ref}),\ \forall i \in\textbf{N}$. Vehicles are also randomly given \textit{a priori} paths, which can either be straight or right-turning. 
	We define $\mu$ as the traffic density, which is the average amount of cars in the simulation environment. The speed ratio of vehicle $i\in\mathbf{N}$ at instant $k\in\mathbb{N}$ is defined as
	\begin{equation}
		\nu_{i}(k)=\frac{v_i(k)}{v_{r_i}}.
	\end{equation}		
	Each vehicle drives the same distance during the simulation, but with a different trajectory. It is of interest to study the speed ratio of every vehicle as function of the distance $p$, in order to highlight eventual critical points where vehicles show common behaviors. 
	In Fig. \ref{fig:speed_ratio_MD}, a simulation is run with a traffic density $\mu=5.15$. Speed ratios $\nu_i(k)$, $\forall i \in\mathbf{N}$, are plotted as function of $p$. Many vehicles present a speed drop in proximity of two critical points, which are regarded as the collision points they are going to cross (subtracted by $\lambda_3$, whose meaning is showed in (\ref{eq:definition_dsafe})). Most likely, vehicles slow down in this position in order to wait for higher priority vehicles to pass. 
	The influence of frontal vehicles can also be discerned from the figure; in fact, even after the collision points, some vehicles keep a constant $v_{i}$, constrained by the frontal vehicle's speed. When this exits the simulation environment, $v_i$ starts seeking the reference speed.
	The following quantity, referred to as the traffic speed ratio, is
	\begin{equation}
		\bar{\nu}(p)=\frac{\sum_{i\in\mathbf{N}}\nu_i(p)}{|\mathbf{N}|},
	\end{equation}
	which is of interest to study traffic properties, since it shows the traffic average speed ratio as function of the position in the intersection. 
	The traffic speed ratio in Fig. \ref{fig:speed_ratio_MD} shows two different behaviors, based on whether a right-turning or a straight-going traffic is considered. In fact, right-turning traffic slows down only before its unique collision point, $\textbf{h}_{j_1}$. Conversely, straight-driving traffic presents a larger speed drop, since it has to cross also the second collision point, $\textbf{h}_{j_2}$.
	Traffic behavior varies if subjected to a higher traffic density, as in Fig. \ref{fig:speed_ratio_HD}, where $\mu=6.31$. Due to the higher number of vehicles involved, the average speed is lower than in Fig. \ref{fig:speed_ratio_MD}, and shows even bigger speed drops in correspondence of the two collision points. 
	
	
	
	
	We refer to \Cref{tab:problemData} for the values used throughout the simulation. 
	\section{Conclusion and future work}
	\label{sec:conclusion}
	We have introduced a decentralized consensus-based control strategy seeking to automate a road intersection. With a consensus-based auction algorithm, vehicles agree on crossing priorities, based on their dynamics. Then, with an on-board MPC controller, every car computes a trajectory, aiming at minimizing a performance measure and avoiding any possible collision with others. Future work will include vehicle models of higher complexity and unpredictable events (like pedestrians crossing in a stochastic way), which will affect the decentralized priority negotiation. In addition, car dimensionality can be modeled with polytopes and, as a consequence, this can be implemented in the collision avoidance system, as in \cite{molinari2017efficient}. Furthermore, concerning the consensus-based auction algorithm, eventual communication failures will be taken into account, as well as different communication topologies, whose convergence speed will be a matter of study. The impact of communication rates on the overall system stability will also be studied. Finally, we aim at dropping the restriction of forbidding left-turning.

	\bibliographystyle{IEEEtran}
	\bibliography{bibliography}
	
\end{document}